# EXOTIC ION-BEAMS, TARGETS AND SOURCES

J.A. Lettry, CERN, Geneva, Switzerland

*Abstract*

Exotic beams of short-lived radioisotopes are produced in nuclear reactions such as thermal neutron induced fission, target or projectile fragmentation and fusion reactions. For a given radioactive ion beam (RIB), different production modes are in competition. For each of them the cross section, the intensity of the projectile beam and the target thickness define an upper production rate. The final yield relies on the optimisation of the ion-source, which should be fast and highly efficient in view of the limited production cross section, and on obtaining a minimum diffusion time out of the target matrix or fragment catcher to reduce decay losses. Eventually, either chemical or isobaric selectivity is needed to confine unwanted elements near to the production site.

These considerations are discussed for pulsed or dc-driven RIB facilities and the solutions to some of the technical challenges will be illustrated by examples of currently produced near-drip-line elements.

## 1 INTRODUCTION

Over the last 30 years, RIB facilities based on Isotope On Line (ISOL) and in-flight fragment separators [1-4] played a major role in very different research domains such as astrophysics, atomic physics, nuclear physics, solid state physics and nuclear medicine.

Nuclei far from stability (so called exotic nuclei) are produced in spallation, fragmentation, fission and fusion nuclear reactions between the primary beam particles and the target nucleus. RIB facilities consist of four parts. The *driver* produces a primary beam of particles and directs them to the *target* where the nuclear reactions take place. This is followed in the case of ISOL by the *ion source* and the *separator* which ionise and select the isotopes and direct them to the physics experiment. The drivers at the origin of the nuclear reaction are based on light or heavy ion accelerators and on various actinide fission processes. The fission of uranium produced via high-energy protons or thermal neutrons is now being complemented by fast neutron induced fission [5] and even photofission [6].

Existing RIB facilities can be classified according to the typical time span between the nuclear reaction and the delivery of the radioisotope to the experimental set-up into three complementary categories: the $\mu s$-class regrouping the in-flight fragment separators; the *ms*-class where fragments are caught and thermalized in a noble gas beam stopper; and the *s*-class where fragments or fission products are stopped or produced at rest in solids, diffuse out of the target material and eventually get ionised in the ion source. Ion sources will be described is section 2 and targets in section 3.

## 2 ION-SOURCES ALONG MENDELEEV'S TABLE

RIB ion-sources should be *fast* to limit the decay losses of short-lived isotopes, *efficient* as the available amount of material is limited by the production cross section, and *chemically selective* to reduce isobar contaminants and confine unwanted species as close as possible to the production area. In this section, a selection of single charge ion sources is briefly presented according to their chemical properties, all of them designed to work close to the high-radiation target area.

### 2.1 Alkalis, alkaline earth and rare earth

A feature common to alkalis, alkaline earth and rare earth elements is their low ionisation potential (between 3.9 and 6 eV). This property, shared by some diatomic molecules, allows a very efficient ionisation on a high temperature metallic surface. The ionisation of an atom bouncing from a high-temperature metallic surface was observed for the first time in 1923. Its efficiency is described as a function of the difference between the energy required to remove an electron from its surface (its work function) and the ionisation potential of the atom. Since then, high-temperature cavities were designed to optimise the ion fraction caught by the extraction field. A commonly used geometry consists of a simple refractory metal (niobium, tantalum, tungsten or rhenium) tube heated by a dc-current.

### 2.2 Metals

Plasma ion sources are based on a controlled electron beam of a few hundred eV which generates a plasma in a low pressure of noble gas in a high-temperature cavity with a low magnetic field. The Forced Electron Beam Arc Discharge (FEBIAD) sources [6-7] efficiently ionise all elements which do not have strong chemical reactions with the ion source materials. Their intrinsic low chemical selectivity is improved via thermo-chromatographic methods. The material and the temperature of the transfer line between the target and the ion source is chosen according to the elements to be condensed.

The Resonant Ionisation Laser Ion Source (RILIS) [8,9] is based on the stepwise excitation of 2 or 3 atomic transitions leading either to auto-ionising states or directly to the continuum. The interaction between the laser beams and the atoms takes place in a high-temperature metallic cavity. While drifting towards the extraction hole, the RILIS ions are confined by a radial potential well originating from the electron layer at the surface of the hot metal. The RILIS elegantly solves the selectivity issue with the notable exception for the alkalis, which are ionised in the high-temperature cavity. It is a proven technique for the efficient production of radioisotopes of 19 elements. In principle an individual ionisation scheme can be developed for 75% of the elements [10]. The RILIS atoms are ionised within the few ns laser pulse, but the different path lengths of drift towards extraction leads to a RILIS ion bunch of typically a few tens of μs duration. The micro-gating technique consists in deflecting the surface-ionised ions from the experimental set-up to a beam dump between two RILIS ion bunches. The selectivity towards surface-ionised isobars can be increased up to a factor of 10 via micro gating [11]. A very elegant solution to the selective ionisation of refractory elements is the Ion Guide Laser Ion Source (IGLIS) [12] at Leuven: recoiling refractory metals fragments are stopped in a noble gas catcher, thermalized and neutralized, and eventually ionised via resonant laser ionization.

*2.3 Halogens*

Halogens have high electron affinities, therefore, a halogen atom hitting a low work function material has a tendency to pick an electron from the surface and to be released as a negative ion. The $LaB_6$ ISOLDE surface ion source [13] efficiently produced negatively charged beams of chlorine, bromine, iodine and astatine. Due to its reactivity, fluorine beams could not be produced this way but rather as a molecular side band ($AlF^+$) in a plasma ion source [14] A dedicated Cs sputter ion sources was tested off-line for fluorine production [15].

*2.4 Gaseous elements*

Despite their high ionisation potentials, noble gases, nitrogen and oxygen are very efficiently ionised in ECR ion sources. Besides the standard high-charge-state ECR, a new generation of ECR designed for single-charge ions was developed [16]. Compact, equipped with permanent magnets and fed with the commercial 2.45 MHz RF generators. They have efficiencies of the order of 90% for $1^+$ argon.

FEBIAD ion sources equipped with a cooled Cu-transfer line are very selective, the only contaminants being the higher charge state of the heavier noble gases. Their efficiency is of the order of 50% for the heavier gases drops to typically 1% for the lighter.

# 3 SELECTED DRIVERS, TARGETS AND ION BEAMS

This section is dedicated to the presentation of examples of targets and drivers dedicated to the production of radioisotopes belonging to specific regions of the chart of nuclides. As more powerful drivers are planned, targets must be designed to cope with higher power densities. The challenge is to dissipate the deposited heat in a target that must be kept at high temperature for maximum diffusion speed. In the cases presented, the radioisotopes are stopped or produced at rest. The release of radioisotopes from thick targets is governed by diffusion out of the target matrix, folded with the usually faster effusion from the surface of the target material to the ion source [17]. A release function gives the probability density for an atom produced at t=0 to be released at a given time [18]. It is the normalised and decay-loss corrected equivalent to the time dependence of the radioisotope current. The sharp rise of the first few ms is usually associated with the effusion process, while the slower tail gives information on the diffusion process. As release curves are often measured with pulsed beams, the induced temperature shock affects the release during the heat transport transient time of typically 1s. The last part of the release function is representative of the diffusion process.

Neutron-rich isotopes are mainly produced in the fission of actinides to benefit from the neutron excess. Fission can be induced by high-energy charged ions, thermal or fast neutrons and even photons [19]. The distribution of the fission fragments in the chart of nuclides depends on the energy of the primary particle, and is widest for high-energy primary particles. In this case, fragmentation and spallation reactions contribute to the production of both light respectively heavy radioisotopes, on top of the fission products. Among the highest cross-sections, the fission of $^{235}U$ via thermal neutrons is characterised by its well-known double peak. On the other side of the valley of stability, isotopes close to the proton drip line have to be produced by spallation of a slightly heavier element, or by fragmentation.

*3.1 A target for a high flux reactor*

Thermal neutron induced fission is, thanks to its very high cross section, a very powerful tool for the production of fission fragments as demonstrated for decades by the OSIRIS facility in Studsvik. The next generation of such facilities was proposed at Grenoble ( PIAFE project at the ILL high flux reactor [20]) and Munich (Munich Accelerator for Fission Fragments MAFF [21] currently under construction). One of the objectives of MAAF is the production of superheavy elements via fusion reaction between two different neutron-rich fission products. This would be realised by implanting neutron-rich fission fragments on line on a substrate, to constantly regenerate a thin very neutron rich target and, simultaneously, accelerate other fission fragments at high energy onto this target [22]. The $^{235}$U primary target consists of 1g uranium carbide in a porous graphite matrix. It is confined in a rhenium vessel directly connected to the ion source (surface, plasma or laser). The lifetime of the target and ion source system has to exceed the cycle of the reactor of 52 days. One of the preferred ion sources to match this challenge is the RILIS, as all complex parts are out of the highly active region except for the high-temperature metallic cavity.

The neutron flux of maximum $3 \times 10^{11}$ n/cm$^2$s can be adjusted via the position of the target at the edge of the reactor. The fission power heats up the target to a nominal temperature above 2000°C. A higher fission rate density can be achieved than with charged particle induced fission schemes where energy losses contribute to the heating. The expected fission rate is $10^{14}$ s$^{-1}$.

### 3.2 Dc vs. pulsed proton beam drivers

The ISOLDE facility driver used to be a 600MeV synchrocyclotron (SC) for 2 decades. Following the move of the facility to CERN's Proton Synchrotron Booster (PSB) [23,24], all targets developed for the dc-beam of the SC were tested with low frequency proton pulses at the PSB. Similar targets are currently very successful at TRIUMF [25] with one order of magnitude higher dc-proton current. The effects of the increase of energy density of 3 orders of magnitude between the almost continuous SC and the pulsed PSB proton beam are briefly described in this section.

*Molten metal targets:* The thermal shock generated by a pressure wave during the 2.4 μs proton pulse was sufficient to break the welds of the target container, to generate vapour pressure bursts, and to splash the molten metal into the ion-source [26]. Corrosion and even cavitation-like attacks were observed on a Ta-container for molten lead. The transfer lines between the containers and the ion-sources were equipped with temperature controlled baffle systems, also designed to condense the excess of vapours. This intermediate temperature control allows in some cases to reach higher temperature. The power of a 1 GeV proton beam at an average current of 1 μA heats up a lead target unit up to 800°C and sets the power dissipation limit of the present design, which can be improved by cooling.

*Metal foil targets:* The rapid release from metal targets benefits from the pulsed production of the radioisotopes, which allows additional noise suppression by collecting data only after the proton pulses, according to the release and to the specific half-life. On the other hand, the temperature increase following the proton pulse can reach 600°C and occurs in metals that are often beyond their elasticity domain. This leads to the destruction of the target container and rapid sintering of the target material. Defocusing the proton beam gives a control over the maximum temperature increase and measurements show the expected increase of the rapidly released isotopes induced by the thermal shock.

The synchronous production of isotopes with half-lives of a few ms duration reduces dramatically the data acquisition time fraction. As an example, for $^{14}$Be (4.35 ms) the background is reduced by a factor 200 for a data acquisition time of more than 12 half-lives.

The pulsed generation of radioisotopes in a target without (or with a very reduced) thermal shock is possible for fission products if the fission is generated with pulsed neutrons. A uranium carbide target placed in the vicinity of a tantalum cylinder bombarded with high-energy pulsed protons corresponds to this description and is presently under test at ISOLDE. The principle of the conversion has been demonstrated at Orsay (section 3.3), Gatchina (ISIS) [27] and of course in spallation neutron sources.

### 3.3 Molten Uranium target

Fast neutrons resulting from the interaction of deuterium with a light converter (beryllium, carbon or liquid lithium) do induce $^{238}$U fission. One of the aims of the first proposal by Nolen [5] was to remove the charged particle energy losses from the high temperature uranium target. This concept was partly tested within the PARRNe project [28,29]. The primary targets called neutron converters were built out of low Z material but as demonstrated by the spallation sources, high Z materials including uranium can be envisaged. The RIB production of uranium carbide and molten uranium targets coupled to graphite and beryllium converters were compared at deuteron energies between 15 and 150 MeV. The net gain of neutron production using Be converters was reduced by geometrical effects as the C-converter can be placed closer to the high-temperature target. While the fast release from uranium carbide matrix is well-known, the first measurement from molten uranium confirmed the usually slow (tens of seconds) release behaviour common to all molten metal targets. The challenging container for molten uranium at high temperature was made out of sintered yttria which kept 200g uranium at temperatures up to 1700°C [30] for a few days. Observations of the container after the run showed no damage but a blackening of the yttria (due to oxygen losses).

These tests will be used as a benchmark for the simulation codes necessary to optimise the numerous parameters such as deuteron energy, converter and target thickness and geometry. With this technique applied in SPIRAL-II, a target of typically 3 kg $^{238}$U would be necessary to achieve of the order of $10^{14}$ fissions/s.

### 3.4 Very thin Ta-foil target for neutron halo nuclides

The half-lives of the neutron halo nuclides $^{11}$Li (8.7ms) and $^{14}$Be (4.34ms) are comparable to the typical effusion time from a standard tantalum foil target. Their diffusion time constants are even orders of magnitude larger. Therefore, the target geometry and the diffusion thickness must be optimised to reduce the large decay losses of four orders of magnitude. Instead of the 130 g/cm$^2$ 20 µm Ta-foils contained in a 20 cm long 2 cm diameter oven only 10 g/cm$^2$ 2 µm Ta-foils were deposited in a u-shaped support and oriented along the beam axis inside the oven or target container [31]. This container was connected to a surface ion-source consisting of a 3 mm-diameter 30 mm long tungsten cavity. While lithium was ionised on the 2400°C surface, the RILIS laser beams ionised the beryllium isotopes. The specific yield from $^{11}$Li increased by a factor of more than 20 and its absolute yield reached 7000 ions/µC (1.4 GeV protons). The best yield of $^{12}$Be (23.6 ms) obtained increased by an order of magnitude, and doubled for $^{14}$Be. This successful example demonstrates that a target has possibly to be designed for each or a few specific isotopes.

### 3.5 Radiation-cooled high-power targets

Radiation-cooled targets or very refractory materials were first proposed by Bennet et al. [32] within the RIST project. The RIST target consists of a succession of diffusion-bonded 25 µm Ta-spacers and discs. A hole of decreasing diameter is cut in the centre to distribute the power along the axis. The target is designed to work with 100 µA pulsed beam of 800 MeV protons. The total electrical power that could be dissipated by radiation at 2400°C in an off-line test reached 30 kW.

The heavy ion beam from GANIL foreseen for the SPIRAL project, has a total power of 6kW. The short range of heavy ions in matter sets stringent constraints on the target/catcher material. The shape of the graphite target developed for noble gas isotopes is a succession of increasing diameter 0.5 mm thick disks held on a central rod which can also be used as heating resistor. The heavy-ion beam circles around the axis. The target is contained in a water-cooled box and the gaps between two adjacent disks are designed to increase the heat-radiating surface (2500°C). In addition they allow a rapid effusion towards the ECR ion source. The micrometer-size graphite structure is optimum for fast diffusion. During the test of the first prototype, the sublimation temperature of graphite was reached and resulted in a hole located at the Bragg peak. A heat transfer code including heat radiation, was developed to define the final geometry of a target which was then able to dissipate the power.

A complementary approach based on conductive cooling was tested at TRIUMF, where a water-cooled set of diffusion-bonded disks of molybdenum representing a medium temperature target was submitted to a 100 µA beam of 500 MeV protons. The heat transfer calculations showed small discrepancies with respect to the recorded temperatures [33].

### 3.6 Target for alkali suppression

The on-line production of neutron-rich copper isotopes is very efficient with high-energy protons impinging on a thorium (or uranium) target. Unfortunately, the yield of rubidium isobars is four orders of magnitude higher. A target unit was therefore designed to purify the copper signal from its rubidium isobars. The micro-gating technique around the typically 30 µs laser ionised Cu bunches contributes to the selectivity by a factor of 4. Thanks to their low ionisation potential, rubidium isotopes will be in a positive charge state already inside the target oven. The dc heated high-temperature tantalum oven generates an electric field, which directs rubidium ions opposite to the RILIS cavity (while the neutral copper atoms are insensitive to this electric field). At this point, a small aperture extraction system collects them. The selectivity gain measured on a Ta-foil target is 5 for rubidium and caesium, and it rapidly drops for elements with higher ionisation potentials. The last, but not least, way to reduce the production of rubidium is to generate the fission via fast neutrons produced in a converter placed close to the target oven. The expected cross-section ratios contribute to the selectivity up to an order of magnitude. These suppression systems are in addition to the selectivity provided by the resolving power of the mass separator.

### 3.7 Ion beam manipulation

The characteristics of radioactive ion beams such as emittance, energy spread or charge state sometimes have to be matched to the physics set-up or to the acceptance of the following beam optics element. For this purposes an increasing number of facilities make use of ion traps and gas cells. An ion trap acting as buncher and used to improve the beam emittance is installed in the REX-ISOLDE LINAC [34] which requires a q/m larger than 0.25. Its charge state breeding relies on the trapping and bunching of the ions prior to the injection into an Electron Beam Ion Source (EBIS). In this trap, up to $10^7$ ions are accumulated and bunched, and their emittance is reduced by one order of magnitude to match the acceptance and the time structure of the EBIS. At the Ion Guide Isotope Separator Leuven (IGISOL) facility, laser ionised refractory metal fragments are transported via a noble gas flow. The skimmer electrode used to extract the singly charged fragments induces an energy straggling which vanished when the skimmer was replaced by a Sextupole Ion Guide (SPIG).

The Rare Isotope Separator (RIA) includes, beside the standard ISOL and in-flight separation, a new production scheme. In this variant, the fragments (possibly produced on a liquid Li target) will be stopped in a noble gas cell where, after thermalization, the singly charged ions are transported by weak electric fields towards the ion beam transport system. Preliminary tests showed that the release time from such a gas cell is well below 1 s [35] depending on the thickness of the gas cell (in this case length × pressure). This part of the RIA project aims at combining the ion beam quality from an ISOL technique and the short release delays of a fragment separator.

## 4 CONCLUSION

The standard complementary methods used in RIB facilities are now completed with an intermediate one, namely the stopping of fragments in gas cells as in the RIA project. The increase of the total beam power available at TRIUMF, or expected for the SIRIUS project, has triggered a large effort in the development of high power targets for high-energy proton driver schemes. The targets developed for moderate intensities or dc. beams could be adapted for beam intensities up to 20 μA and for the today's pulsed drivers. The limitation on the maximum power which can be deposited in a small target volume, can be overcome partially for the production of fission products, by the introduction of a converter producing fast neutrons. However, fission in reactor-based facilities like MAFF sets a challenging goal of $10^{14}$ fissions/s.

A large R&D effort from the whole RIB community is visible, as witnessed by the RIA project and by the creation of study groups on targets, traps and ion-sources, triggered by the EURISOL facility project. Obviously, with the expected intensities, not only the target area, but in a less critical manner the beam transport, traps and experimental set-ups will have to be designed with regard of their contamination when handling the very intense beams foreseeable in the near future.

## REFERENCES

[1] H. Grunder, Proc. of the RNB–5 Conf. Divonne 2000, to be published in Nucl. Phys. A.
[2] B. Jonson, Proc. of the RNB–5 Conf. Divonne 2000, to be published in Nucl. Phys. A.
[3] I. Tanihata, ENAM 98 AIP Conf. Proc. **455**, 943 (1998).
[4] H. Ravn, Radioactive ion-beam projects based on the two accelerator or ISOL principle, Phil. Trans. R. Soc. Lond. A **365**, 1955 (1998).
[5] J. Nolen, RNB-3 Conf. Proc. D. Morissey Ed., Editions Frontières, Gif sur Yvette. 111, (1993).
[6] R. Kirchner and E. Roeckl, Nucl. Instrum. and Meth. **133**, 187 (1976).
[7] R. Kirchner, K. H. Burkard, W. Hüller and O. Klepper, Nucl. Instrum. and Meth. **186**, 295 (1981).
[8] V.I. Mishin et al., Nucl. Instrum. and Meth. B **73**, 550 (1993).
[9] A. E. Barzakh, et al., Nucl. Instrum. and Meth. B **126**, 85 (1997).
[10] U. Koester et al., Proc. of the RNB–5 Conf. Divonne 2000, to be published in Nucl. Phys. A.
[11] J. Lettry et al., Rev. Sci. Instrum., **69**, 761 (1997).
[12] Y. Kudriavtsev et al., Nucl. Instrum. and Meth. B **114**, 350 (1996).
[13] B. Vosicki et al., Nucl. Instrum. and Meth. **186**, 307 (1981).
[14] R. Welton et al., Proc. of the RNB–5 Conf. Divonne 2000, to be published in Nucl. Phys. A.
[15] G. Alton, et al., Nucl. Instrum. and Meth. B **142**, 578 (1998).
[16] A. Villari et al., Proc. of the RNB–5 Conf. Divonne 2000, to be published in Nucl. Phys. A.
[17] J.R.R. Bennett, Proc. of the RNB–5 Conf. Divonne 2000, to be published in Nucl. Phys. A.
[18] J. Lettry et al., Nucl. Instrum. and Meth. **B** 126 (1997) 130.
[19] Y. Oganesian, Proc. of the RNB–5 Conf. Divonne 2000, to be published in Nucl. Phys. A.
[20] PIAFE project report, ed. By U. Köster and J.-A. Pinston, ISN Grenoble, 1998.
[21] D. Habs et al., Nucl. Phys. A616 (1997). http://www.ha.physik.uni-muenchen.de/maff/
[22] U. Köster, AIP Conf. Proc. 475, 269, (1998).
[23] B. Allardyce and R. Billinge, CERN PS/DL 89-37.
[24] B. Allardyce et al., CERN PS/92-46 (PA).
[25] R. Laxdal et al., These proceedings.
[26] J. Lettry et al., Nucl. Instrum. and Meth. **B** 126 171, (1997).
[27] J. Nolen and V. Panteleev, communication to the TWIST workshop GANIL Caen 2000.
[28] S. Kandri-Rody et al., Nucl. Instrum. and Meth. B160, 1, (2000).
[29] F. Clapier et. al., Phys. Rev. Special topics, Accelerators and Beams, 1, 013501 (1988).
[30] O. Bajeat, communication to the TWIST workshop GANIL Caen 2000.
[31] J.R.J. Bennett et al., Proc. of the RNB–5 Conf. Divonne 2000, to be published in Nucl. Phys. A.
[32] J.R.J. Bennett, Nucl. Instrum. and Meth. B **126,** 105 (1997).
[33] W. Talbert et al., Proc. of the RNB–5 Conf. Divonne 2000, to be published in Nucl. Phys. A.
[34] O. Kester et al., these proceedings.
[35] G. Savard et al., Proc. of the RNB–5 Conf. Divonne 2000, to be published in Nucl. Phys. A.